\documentstyle[prd,aps,psfig,floats]{revtex}

\def\etal{et al.}

\def\kmsmpc{km~s$^{-1}$~Mpc$^{-1}$}

\def \ie {i.e.}
\def \lleq {\lower0.9ex\hbox{ $\buildrel < \over \sim$} ~}
\def \ggeq {\lower0.9ex\hbox{ $\buildrel > \over \sim$} ~}
\def \rb{\bar{r}}
\def \sb{\bar{s}}
\def\l{\Lambda}

\def\spose#1{\hbox to 0pt{#1\hss}}
\def\simle{\mathrel{\spose{\lower 3pt\hbox{$\mathchar"218$}}
     \raise 2.0pt\hbox{$\mathchar"13C$}}}
\def\simge{\mathrel{\spose{\lower 3pt\hbox{$\mathchar"218$}}
     \raise 2.0pt\hbox{$\mathchar"13E$}}}
\def\apj{Astroph.~J.~}
\def\mn{Mon.~Not. Roy.~ast. Soc.~}

\def\aj{Astron.~J.~}
\def\prl{Phys.~Rev. Lett.~}
\def\prd{Phys.~Rev.~D~}

\def\plb{Phys.~Lett.~B~}

\def\beq{\begin{equation}}
\def\eeq{\end{equation}}
\def\ber{\begin{eqnarray}}
\def\eer{\end{eqnarray}}

\newcommand{\sq}{\lower.25ex\hbox{\large$\Box$}}

\tighten

\begin{document}

\draft
\preprint{IUCAA-41/99}
%\twocolumn[
%\hsize\textwidth\columnwidth\hsize\csname@twocolumnfalse\endcsname
\title{Statefinder -- a new geometrical diagnostic of dark energy}
\author{
Varun Sahni${}^{a}$, Tarun Deep Saini${}^{a}$,
A. A. Starobinsky${}^{b}$ and Ujjaini Alam ${}^{a}$}

\address{${}^{a}$ Inter-University Centre for Astronomy \& Astrophysics,
Pun\'e 411 007, India}
\address{${}^{b}$ Landau Institute for Theoretical Physics,
119334 Moscow, Russia}
\date{\today}
\maketitle
\begin{abstract}
%In order to characterize the properties of dark energy 
%in a model independent manner and 
%from a purely geometrical point of view, 
We introduce a 
new cosmological diagnostic pair $\lbrace r,s\rbrace$ called Statefinder. 
The Statefinder is a geometrical diagnostic and allows us to 
characterize the properties of dark energy
in a model independent manner.
The Statefinder is dimensionless and is constructed from the scale factor 
of the Universe and its time derivatives only. The parameter $r$ forms 
the next step in the hierarchy of geometrical cosmological parameters 
after the Hubble parameter $H$ and the deceleration parameter $q$, while 
$s$ is a linear combination of $q$ and $r$ chosen in such a way that it 
does not depend upon the dark energy density. The Statefinder pair 
$\lbrace r,s\rbrace$ is algebraically related to the
equation of state of dark energy and its first time derivative. 
The Statefinder pair is 
calculated for a number of existing models of dark energy having both 
constant 
and variable $w$. For the case of a cosmological constant the Statefinder
acquires a particularly simple form. 
We demonstrate that the Statefinder diagnostic can 
effectively differentiate between different forms of dark energy. 
We also show that the 
mean Statefinder pair can be determined to very high accuracy from 
a SNAP-type experiment.

\end{abstract}

% for PACS codes, see http://publish.aps.org/PACS/pacs99.html
 %Mathematical and relativistic aspects of cosmology;
% 97.60.Bw} %Supernovae
\bigskip
%]\renewcommand{\thefootnote}{\arabic{footnote}} \setcounter{footnote}{0}
%\narrowtext
\widetext

Recent observations of type Ia supernovae indicate that the expansion
of the Universe is accelerating rather than slowing down 
\cite{perlmutter99}. These results, when combined with cosmic 
microwave background (CMB) observations of a peak in the angular power 
spectrum on degree scales \cite{cmb,benoit}, strongly suggest that the 
Universe is spatially flat with $\sim 1/3$ of the critical energy density 
being in non-relativistic matter and $\sim 2/3$ in a smooth component 
with large negative pressure (`dark energy'). Indirect support for dark 
energy (known long ago) comes from the examination of gravitational 
clustering within the framework of the standard gravitational instability 
scenario (see the reviews \cite{ss99,p01}). Finally, with recent data on
the galaxy power spectrum from 2dF Galaxy Survey combined with CMB data, 
the existence of dark energy can be proved without using the supernovae 
data at all \cite{perc}. A large body of recent work has focussed on 
understanding the nature of dark energy and its possible relation to a 
fundamental theory of matter such as M-theory, supergravity etc. Despite 
the considerable effort in this direction, both the nature of dark energy 
as well as its cosmological origin remain enigmatic at present.

The simplest model for dark energy is a cosmological constant
$\l$, whose energy density remains constant with time $\varepsilon = 
\l/8\pi G$ and whose effective equation of state remains fixed, 
$w\equiv P/\varepsilon = -1$ ($P$ is the pressure) as the Universe 
evolves. The cold dark matter (CDM) model with the cosmological constant 
having the corresponding mass density
\beq
\rho_{\l}={\varepsilon_{\l}\over c^2} = 6.44 \times 10^{-30}
\left({\Omega_{\l}\over 0.7}\right)\left({h\over 0.7}\right)^2 
{\rm {g~cm^{-3}}}~,
\label{Lambda}
\eeq
where $h$ is the Hubble constant $H_0$ in terms of $100$ \kmsmpc 
and $\Omega_{\l}=0.7\pm 0.1,~h= 0.7\pm 0.1$, provides an excellent 
explanation for the acceleration of the Universe and other existing 
observational data. However, it remains quite possible that the dark 
energy density may depend sufficiently weakly upon time. This follows
from many proposed models. The possibility that dark energy could be 
dynamical is also suggested by 
the remarkable {\em qualitative} analogy between the observed properties 
of dark energy and properties of a different type of `dark 
energy' -- namely the inflaton field -- 
postulated in the inflationary scenario 
of the early Universe.
 
%These properties of $\Lambda$ result in a fine tuning problem:
%the relative densities in the cosmological constant
%and in radiation must be set to an accuracy of better than one
%part in $10^{123}$ at the Planck time, in order to ensure
%that the densities in matter
%and the cosmological constant become comparable at {\em precisely}
%the present epoch.

Once we allow the dark energy density to be time-dependent, then the
next simplest class of models are those with a {\em constant}, 
non-positive $w$. We shall call this class
`quiessence' (Q) ($w<-1/3$ is a necessary condition to make the universe
accelerate). Examples include a tangled and 
`frustrated' network of cosmic strings $w = -1/3$ and domain walls 
$w = -2/3$. More generally, in a Friedmann-Robertson-Walker (FRW)
background with the presence of CDM, an arbitrary but constant 
$w$ for dark energy 
from the range $(-1,0)$ can be achieved by using a scalar field with
a hyperbolic sine potential (see Eq. (\ref{quequi}) below)\cite{ss99}. 
It may be noted that in principle the value of 
$w$ may be even less than $-1$; the present 
observational data do not exclude this possibility but limit the 
constant $w$ in the range of about $(-1.6,-0.8)$\cite{mel02}. 

A more generic alternative to $\Lambda$ and Q is presented by `kinessence' 
(K) which refers to dark energy with a {\em time dependent} $w$. Examples 
of kinessence include `quintessence' -- a scalar field $\phi$ with a 
self-interaction potential $V(\phi)$ minimally coupled to gravity (see
\cite{ss99} for numerous references), as well as the `Chaplygin gas'
model \cite{kam01} and braneworld models of
dark energy \cite{dgp00,ssh02}. These three alternatives are 
summarized in the Table I (where $z \equiv a(t_0)/a(t) -1$ is the
redshift, $a(t)$ is a FRW scale factor and the subscript $0$
denotes the present moment). 

%`Tracker' potentials lead to  an effective scalar field equation of 
%state which remains close to the background value during most of 
%the expansion history of the Universe. Close to the present epoch the 
%equation of state of Kinessence turns negative, and K begins to dominate 
%the energy density of the Universe and drive its current accelerated 
%expansion. The fact that this has happened very recently (at $z<1$) is 
%sometimes referred to as the `cosmic coincidence problem'.

\begin{table*}[tbh!]
\begin{center}
\begin{minipage}[h]{0.9\linewidth} \mbox{} \vskip -18pt
\bigskip
\begin{tabular}{lll} %\hline
Dark Energy& State Parameter & Energy Density Parameter\\\hline
 Cosmological constant         &   $w(z) = {\rm constant} = -1$  &
 $\rho(z) = \Lambda/8\pi G = {\rm constant}$\\
 Quiessence          &   $w(z) = {\rm constant} < -1/3$ &
$\varepsilon(z) = \varepsilon_0(1+z)^{3(1+w)}$\\
Kinessence & $w(z) \neq {\rm constant} $ &
$\varepsilon(z) = \varepsilon_0\exp{\left [3\int_0^z dz' \frac{1+w(z')}
{1+z'} \right]}$\\
\end{tabular}
\label{table:dark}
\end{minipage}
\end{center}
\medskip
\caption{
}
\end{table*}

\bigskip

The effective equation of state is clearly an important
property of dark energy. This has led to numerous attempts to reconstruct 
the former 
from observations of high redshift supernovae in a model independent 
manner \cite{star98,srss00,reconst}. However, for field-theoretical 
models of dark energy, the equation of state is not a {\em fundamental} 
property. Strictly speaking it has reference only
to an exactly isotropic FRW background. For 
small perturbations superimposed on a FRW background, the pressure 
tensor is generically non-diagonal (non-barotropic), and the velocity 
of signal propagation need not be given by the standard hydrodynamic 
expression $\sqrt{dP/d\varepsilon}$. Moreover, the very notions of 
$\varepsilon$ and $P$ for dark energy pre-suppose the {\em Einstein 
interpretation} of gravitational filed equations (not to be confused with 
the notion of the Einstein frame which is used in scalar-tensor and 
string theories of gravity!). Namely, even if the real equations for a 
given model are not the 4-D Einstein equations at all (examples include
dark energy models in scalar-tensor \cite{beps00} and brane 
\cite{dgp00,ssh02} gravity), one can still write them formally
in the Einstein 
form, by placing the Einstein tensor $R_{ij}-{1\over 2}g_{ij}R$ 
into the left-hand side, and by
grouping all other terms in the right-hand side
and calling them (after dividing by $8\pi G$) `the effective 
energy-momentum tensor of matter'. After that, the energy-momentum tensor 
of dust-like matter (describing CDM and baryons) is subtracted from 
the latter, and the remaining part is used to define $\varepsilon$ and 
$P$ for `dark energy'. All this reveals how ambiguous the notion of
`equation of state' can be for a non-Einsteinian model of dark energy.

Fundamental variables (at least, at the field-theoretical level of 
consideration) are either geometrical (astronomical) -- if they are 
constructed from a space-time metric directly, or physical -- those 
which depend upon properties of physical fields carrying dark energy. 
Physical variables are, of course, model-dependent, while geometrical 
variables are more universal. Additionally, the latter do not 
depend upon uncertainly measured physical quantities such as the present 
density of dust-like matter $\Omega_m$. That is why we emphasise the use 
of geometrical variables when describing the present
expansion of the Universe and properties of `dark energy'.

The oldest and most well-known geometric variables are the Hubble 
constant $H_0$ and the current value of the
deceleration parameter $q_0$. At present, 
accurate measurements of the expansion law of the Universe during the past
are also possible (e.g., using the luminosity distance to
distant supernovae), therefore these variables should be generalized to
the Hubble parameter $H(t)\equiv \dot a/a$ and the deceleration
parameter $q(t)\equiv -a\ddot a/\dot a^2= -\ddot a/aH^2$ 
($H_0=H(t_0)$ and $q_0=q(t_0)$). However, both the necessity of 
consideration of more general models of dark energy than a cosmological 
constant, and the remarkable increase in the accuracy of cosmological 
observational data during the last few years, compel us to advance 
beyond these two important quantities. 
For this reason, in this letter we propose 
a new geometrical diagnostic pair for dark energy. This 
diagnostic is constructed from the $a(t)$ and its derivatives up to the 
third order. Namely, we introduce the {\em Statefinder} pair 
$\lbrace r,s\rbrace$:
\beq
r = \frac{\stackrel{...}{a}}{a H^3}, ~~s = \frac{r - 1}{3\,(q - 1/2)}.
\label{eq:state}
\eeq
$r(z)$ is a natural next step beyond $H(z)$ and $q(z)$. We will soon 
see that it has a remarkable property for the basic flat $\l$CDM FRW
cosmological model. $s(z)$ is a linear combination of $r(z)$ and
$q(z)$. In a companion paper we shall show that a particular
combination of two variables from the above three e.g., $q$
and $s$, can provide an excellent diagnostic 
for describing the properties of dark energy \cite{alam}.

Below we will assume that the Universe is spatially flat, $k=0$. This
assumption naturally follows from the simplest versions of the 
inflationary scenario and is convincingly confirmed by recent CMB 
experiments \cite{benoit}. At late times ($z \lleq 10^4$) the Universe is 
well described by a two component fluid consisting of non-relativistic
matter (CDM+baryons) $\Omega_m$ and dark energy $\Omega_X=1-\Omega_m$. In 
this case the Statefinder pair acquires the form
\ber
r &=& 1 + \frac{9}{2}\Omega_X w(1+w)
- \frac{3}{2}\Omega_X \frac{\dot w}{H}, \label{eq:state1}\\
s &=& 1 + w - \frac{1}{3}\frac{\dot w}{w H}
\label{eq:state2}
\eer
where $w=P_X/\varepsilon_X$. Thus, if the role of dark energy is played 
by a cosmological constant ($w=-1)$, then the value of $r$ stays pegged 
at $r=1$ throughout the {\em entire matter dominated epoch and at all 
future times}; \ie , $r\equiv 1$ for $z \lleq 10^4$ irrespective of the 
current value of $\Omega_m$. The extreme simplicity of the parameter 
$r(z)$ for the basic cosmological model ($\Lambda$CDM)
which also provides the best fit to 
existing observational data may, in fact, prove not to be a mere
coincidence !
\footnote {Note that the quantity $r(z)$ was also considered in 
the paper \cite{cn98} for a non-flat case when it is time-dependent. 
However, its remarkable property for the flat $\Lambda$CDM model was 
not emphasized. For completeness, let us mention that $r=2q=\Omega_m(z),~
s\equiv 2/3$ for a matter dominated 
non-flat CDM model with negligible amounts of dark energy and radiation.}
Very different behaviour is predicted for quiessence 
and kinessence for which $r$ is a function of time. In particular,
if dark energy is attributed to a minimally coupled scalar field $\phi$
(quintessence),
\beq \label{r-phi}
r= 1+{12\pi G\dot\phi^2\over H^2} + {8\pi G{\dot V}\over H^3}~.
\eeq

%For quiessence models ${\dot w} = 0$, 
%whereas many Kinessence models, including those with the `tracker' 
%stage in the past, have ${\dot w} < 0$. For such models,  
%\footnote {However, ${\dot w}$ may well be positive even if dark energy 
%is attributed to the energy of a minimally coupled scalar field; the 
%simplest example is a massive scalar field, $V(\phi)=m^2\phi^2/2$. In 
%such models, the epoch of dark energy domination is typically 
%{\em transient}.} As a result, the value of $r$ in K \& Q models will 
%monotonically decrease as the Universe expands and evolves. The current 
%value $r_0$ should therefore help establish whether dark energy is a 
%cosmological constant $(r_0 = 1)$ or has some other origin 
%$(r_0 \not= 1)$.

The properties of the second Statefinder `$s$' complement those of the 
first. For the basic $\l$CDM model with any non-zero $\l$, $s\equiv 0$. 
Moreover, $s$ does not depend neither on time, nor on $\Omega_m$,
for quiessence models for which $s = 1+w$. In marked contrast, $s$ 
generically depends on time for kinessence. E.g., for quintessence:
\beq\label{s-phi}
s={2\left(\dot\phi^2 +{2\dot V\over 3H}\right)\over \dot\phi^2-2V}~.
\eeq 
Thus, properties of the Statefinder pair $\lbrace r, s\rbrace$ enable it 
to differentiate between the three canonical forms of dark energy 
described in Table 1.

%In this context it is worth recalling that the basic observational 
%parameters in cosmology are geometrical in nature, since they involve 
%either differential or integral quantities constructed out of the scale 
%factor. The Hubble and deceleration parameters $H, q$
%and the Statefinder pair $\lbrace r, s\rbrace$
%provide us with examples of the former, while the coordinate distance
%$r_c = \int dt/a$, is an example of the latter. The luminosity distance 
%$d_L$ (which is based on the notion of an absolute candle), and the 
%angular-size distance $d_A$ (based on the standard ruler concept) are 
%both derived from the coordinate distance: $d_L = (1+z) a_0 r_c(z), 
%d_A = (1+z)^{-1} a_0 r_c(z)$. Since geometrical and physical parameters
%are related to one another through the field equations of physical 
%cosmology, a major concern of observational cosmology over the past few 
%decades has been to determine the values of physical parameters on the 
%basis of accurate measurements of geometrical observables.
%If dark energy has the form of a minimally
%coupled scalar field, the physical
%parameters which provide us with the deepest insight include:
%$\Omega_m(t), \Omega_X(t) \equiv \Omega_V(t) + \Omega_T(t) = 
%1-\Omega_m(t),
%\Omega_V(t) =  8\pi G V(t)/3 H^2(t), \Omega_T(t) = 8\pi G T(t)/3 H^2(t), 
%(T = \dot\phi^2/2)$ and $\Omega_\Pi(t) = 8\pi G{\dot V}(t)/3H^3(t).$

It is straightforward to invert Eqs. (\ref{eq:state1},\ref{eq:state2}) 
and express $w$ and $\dot w$ in terms of the Statefinder pair. However, 
$w$ is more directly related to the deceleration parameter:
\beq
w(t)= {2q(t)- 1 \over 3\Omega_X}~.
\eeq
Thus, $w$ a composite quantity since it is constructed out of physical 
($\Omega_X$) as well as geometrical ($q$) parameters. Note that for 
quintessence, $w>-1$ but $\dot w$ may have any sign (for models with 
$\dot w>0$ and $w < 0$ the epoch of dark energy domination is usually 
a {\em transient}). The relationship between geometrical and physical 
parameters is summarized in Table 2.

\begin{table}[tbh!]
\begin{center}
\caption{Relationship between geometrical and physical parameters
characterizing the observable Universe}
\bigskip
\begin{tabular}{ll}
Geometrical parameters & Related physical parameters\\
\tableline
$H = {\dot a}/a$ & $\Omega_{\rm total}, \Omega_{\rm curvature}$\\
$q = -{\ddot a}/aH^2$ & $\Omega_i, w_i$\\
$r = {\stackrel{...}{a}}/{a H^3}$ & $\Omega_i, w_i, {\dot w_i}$\\
$s = (r - 1)/3(q - 1/2)$ & $w_i, {\dot w_i}$\\
%$r_c$ & $\Omega_i, w_i$\\
%\tableline
\end{tabular}
\label{table:lambda}
\end{center}
\end{table}
\bigskip

Let us now study the Statefinder pair for different models of dark energy 
in greater detail. As was mentioned already, its value is equal to
$\lbrace 1,0\rbrace$ for any $\l$CDM model with a non-zero $\l$.
Quiessence models (QCDM) have a constant $w$, as a result
\beq \label{ques}
r = 1 + \frac{9}{2}\Omega_Qw(1+w),~~s = 1+w~.
\eeq
Two values of the equation of state are singled out for special 
attention: $w = -1/3$ (cosmic strings) and $w = -2/3$ (domain walls). 
In both cases the first Statefinder has the simple form $r(t) = 1 - 
\Omega_Q(t)=\Omega_m(t)$. As a result,
$r(t) \to 1$ for $t \ll t_0$, $r(t) \to 0$ for $t \gg t_0$ and
$r_0 \simeq 0.3$ at the present time  when $\Omega_Q(t_0) \simeq 0.7$.
This leads to a degeneracy in $r_0$ for the dual value $w = -1/3, -2/3$.
Though generic, this degeneracy is easily broken when one adds
information from the second Statefinder $s$. Note that the case of an
arbitrary $-1<w<0$ in the presence of a non-zero $\Omega_m$ can be 
achieved using quintessence with the potential \cite{ss99} (see also
\cite{um00}) \footnote{There are some misprints in numerical 
coefficients in Eqs~(119-121) of \cite{ss99} which are corrected here.}
\ber
V(\phi)={3H_0^2(1-w)(1-\Omega_{m0})^{1/|w|} \over 16\pi G\Omega_{m0}^
{(1+w)/|w|}} \sinh^{-2(1+w)/|w|}\left(|w|\sqrt{{6\pi G\over 1+w}}
(\phi-\phi_0+\phi_1)\right)~, \label{quequi}\\
\Omega_{m0}=\Omega_m(t_0),~~\phi_0=\phi(t_0),~~\phi_1= \sqrt{{1+w\over
6\pi G}}{1\over |w|} \ln {1+\sqrt{1-\Omega_{m0}}\over 
\sqrt{\Omega_{m0}}}~. \nonumber
\eer
In this case, $r<1,~0<s<1$.

Let us now turn to the quintessence case where $r$ and $s$ are given 
by Eqs. (\ref{r-phi}) and (\ref{s-phi}) correspondingly. To this 
category belong scalar fields with `tracker' potentials, for which the 
scalar field $\phi$ approaches a common evolutionary path from a wide 
range of initial conditions \cite{track}. Tracker potentials satisfy 
$V''V/(V')^2 \geq 1$. We consider the simplest case of an inverse 
power-law potential $V(\phi) = V_0/\phi^\alpha,~\alpha>0$ first 
studied in \cite{ratra}. For this potential, the region of initial 
conditions for $\phi$ for which the tracker regime has been reached 
before the end of the matter-dominated stage is $\phi_{in}\ll M_P\equiv 
\sqrt G$, and the present value of quintessence is $\phi(t_0)
\sim M_P$. The evolving
values of the Statefinder pair for this potential with 
$\alpha = 2$ and $\alpha=4$ are shown in Figs. 1 and 2. Also shown are 
results for the cosmological constant and quiessence. During 
tracking $\frac{\varepsilon_\phi}{\varepsilon_m} \propto t^{4/(2 + 
\alpha)}$ as a result quintessence always becomes dominant at late times. 
The equation of state of quintessence and the corresponding value of
the Statefinder pair is
given by
\beq
w = -\frac{w_B + 2}{\alpha+2},~~r\approx 1,~~s\approx 1+w
\label{eq:kin_state}
\eeq
($w_B = 1/3,0$ during the radiation- and matter-dominated epochs
respectively). 

%which have played a particularly important role in alleviating the 
%`fine-tuning' problem faced by quiescent models in general and by the 
%cosmological constant in particular.A large class of potentials has been 
%explored for tracker properties. Potentials based on the exponential 
%\cite{exp} and the inverse power-law \cite{ratra} have been particularly 
%successful in producing a late-time accelerating Universe which is in 
%agreement with most recent cosmological observations.
%The equation of motion of the scalar field which plays the
%role of Kinessence is
%\beq
%{\ddot \phi} + 3H {\dot\phi} + V' = 0,
%\eeq
%where
%\beq
%H^2 = \frac{8\pi G}{3} \left(\rho_m + {1\over 2}\dot\phi^2+
%V(\phi)\right),
%\eeq
%$\rho_m=(3H_0^2/8\pi G)\Omega_{0m}(a/a_0)^{-3}$
%includes the density of dust-like `cold' dark matter as well as baryonic 
%matter.Figures 1 \& 2 show results obtained for the tracker field with 

\noindent
\begin{figure}[tbh!]
\centerline{
\psfig{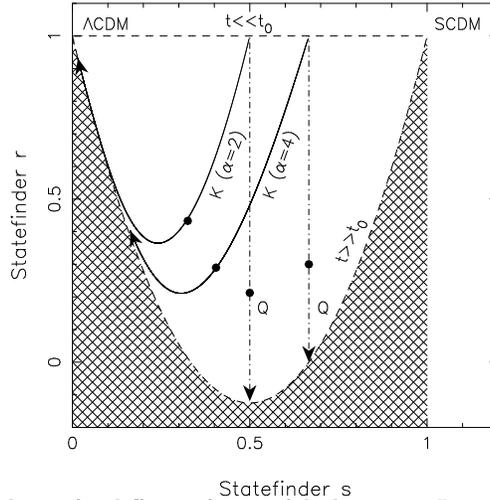} }
\caption{\small The Statefinder pair ($r,s$) is shown for different
forms of dark energy. In quiessence (Q) models ($w = $constant $\neq -1$)
the value of $s$ remains fixed at $s = 1+w$ while the value of $r$
asymptotically declines to
$r(t \gg t_0) \simeq 1 + \frac{9w}{2}(1+w)$.
%which can be obtained
%after substituting ${\dot w} = 0, \omx \to 0$ in (\ref{eq:statefinder}).
Two models of quiessence corresponding to $w_Q = -0.25, -0.5$ are shown.
Kinessence (K) models
are presented by a scalar field (quintessence) rolling down
the potential $V(\phi) \propto \phi^{-\alpha}$ with $\alpha = 2,4$.
These models commence their evolution on a tracker trajectory described 
by (\ref{eq:kin_state}) and
asymptotically approach $\Lambda$CDM at late times.
$\Lambda$CDM ($r=1,s=0$) and SCDM in the absence of $\l$ ($r=1,s=1$) are 
the fixed points of the system.
The hatched region is disallowed in quiessence models and in the
kinessence model which we consider.
The filled circles show the {\em current values} of the Statefinder pair
($r,s$) for the Q and K models ($\Omega_{m0} = 0.3$).
}
\label{fig:state}
\end{figure}

\noindent
\begin{figure}[tbh!]
\centerline{
\psfig{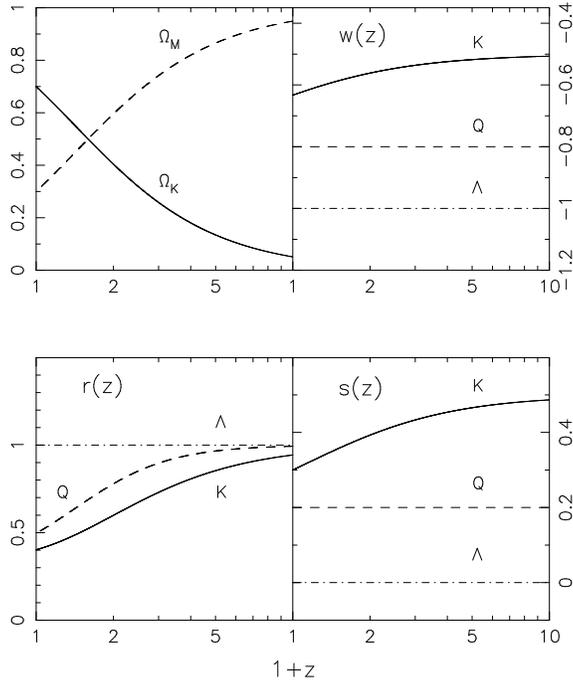} }
\medskip
{\caption{\small The Statefinder pair $\lbrace r,s\rbrace$ is shown
for dark energy consisting of a cosmological constant $\l$,
quiessence `Q' with an unevolving equation of state $w = -0.8$ and
the inverse power law tracker model $V= V_0/\phi^2$,
referred here as kinessence 'K'. The lower left panel shows $r(z)$ 
while the lower right panel shows $s(z)$.
Kinessence has a time-dependent equation of state which is shown in
the top right panel. The fractional density in matter and
kinessence is shown in the top left panel.
}
\label{fig:statefinders1}}
\end{figure}

Constraints from structure formation and the CMB suggest that dark 
energy must be subdominant at $z \ggeq 1$. Primordial nucleosynthesis 
arguments impose the stringent constraint: $\Omega_X < 0.05$ at 
$z \sim 10^9$ \cite{bm01}. Small values of $\Omega_X$ and $w$ 
substantially decrease the terms $\Omega_X w$ and $\Omega_X{\dot w}/H$ 
which appear in the RHS of (\ref{eq:state1}) and ensure that the 
Statefinder $r$ remains close to unity at high $z$. This is exactly 
what one finds from Fig. 2. The extreme sensitivity of $r$ to an
evolving equation of state of the tracker field is reflected by the 
fact that the value of $r$ declines rapidly as the Universe expands, 
dropping to $\sim 50\%$ from its starting value by $z \sim 1$, even 
though dark energy remains subdominant at this epoch. 

%In contrast to the
%time-dependent behaviour of $r$ for Quiessence \& Kinessence,
%the value of $r$ remains fixed at $r=1$ for the cosmological 
%constant $\l$, which makes it easier to discern the latter from 
%Q \& K.Since the the second Statefinder `$s$' distinguishes between 
%Q and K, we conclude that the complementary properties
%of the Statefinder pair $\lbrace r_0,s_0\rbrace$
%make it a useful diagnostic of dark energy.
%Finally $r \to 0$ as $t \to \infty$ in a Kinessence-dominated Universe.
%(The value of $r$ declines from unity, passes through a minimum value
%and rise again towards unity, in
%a Universe in which accelerated expansion
%is a transient epoch occurring between two
%matter dominated stages.)

As is apparent from Fig. 2, the discriminating power of $r$ and $s$ can 
be significant even at moderate redshifts. Since $\Omega_\l$ and 
$\Omega_{\rm Q}$ usually decrease faster with redshift than 
$\Omega_{\rm K}$, the value of $r(z)$ for both the cosmological constant 
and quiessence is generally closer to unity at a given large redshift
than the corresponding value for a tracker field (kinessence).
Thus, whereas the current value of $r_0$ allows us to differentiate
$\l$ from Q and K, the value of $r$ at moderate redshifts distinguishes 
K from  $\l$ and Q. This feature is even more pronounced in the second 
Statefinder $s$, whose value does not explicitly depend upon $\Omega_X$
and whose capacity to distinguish between $\l$ and quiessence on the one 
hand, and from kinessence on the other, actually {\em increases} with 
redshift (see Fig. 2). The present CMB, SNe and galaxy clustering data 
strongly suggest that $\alpha \lleq 1$ for quintessence with the inverse 
power-law potential \cite{mel02}. However, even then the Statefinder 
remains a useful diagnostic as will be shown below.

Let us consider another form of kinessence. Below we determine the value
of the Statefinder pair 
for the simplest of brane cosmological models -- the 
Dvali-Gabadadze-Porrati (DGP) model \cite{dgp00}. It is important to note
that in this model `dark energy' is not the energy associated with a 
new form of matter, rather its origin is geometrical in nature and is
entirely due to the fact that general relativity is formulated in 5 
dimensional space-time. 
The model has only 
one adjustable parameter $r_c$ -- the scale beyond which gravity becomes
five-dimensional. This scale can be related to the current values of 
$H_0$ and $\Omega_{m0}$ by the relation $H_0r_c=1/(1-\Omega_{m0})$. 
The FRW equation for this model reads:
\beq\label{DGP}
H= \sqrt{{8\pi G \varepsilon_m\over 3}+{1\over 4r_c^2}}+{1\over 2r_c}~.
\eeq  
(the choice of sign in front of the last term in the right-hand side 
corresponds to the 'BRANE2' class of models according to the terminology 
of \cite{ssh02}).

The solution to (\ref{DGP})
can be written in the following parametric form:
\ber 
a=a_1\sinh^{2/3}\psi~,~~~{3t\over 2r_c}=\psi+{1-e^{-2\psi}\over 2}~, 
\nonumber \\
H={e^{\psi}\over 2r_c\sinh \psi}~,~~~\varepsilon_m={3\over 32\pi Gr_c^2
\sinh^2\psi}~.~~~\Omega_m\equiv {8\pi G\varepsilon_m\over 3H^2}=
e^{-2\psi}~. \label{brane-sol}
\eer
The values of the deceleration parameter and the Statefinder pair read:
\beq\label{brane-qrs}
q={2\Omega_m-1\over 1+\Omega_m}~,~~~r= 1-{9\Omega_m^2(1-\Omega_m)\over
(1+\Omega_m)^3}~,~~~s={2\Omega_m^2\over (1+\Omega_m)^2}~.
\eeq
In particular, $r=0.74,~s=0.11$ for $\Omega_m=0.3$. At large redshifts
the universe becomes matter dominated and $r\to 1$, $s\to 0.5$.

%The diagnostic properties of the Statefinder may also extend
%to the `mean Statefinder statistic' $\lbrace {\bar r},{\bar s}\rbrace$, 
%${\bar r} = \int_0^1 r(z) dz, {\bar s} = \int_0^1 s(z) dz$.
%
%\section{Determining the Statefinder using SNAP}

At the end of the paper, we estimate the accuracy with which the
Statefinder pair (averaged over a range of $z$) can be determined in 
future SNAP-type satellite missions. The `SuperNovae Acceleration Probe' 
(SNAP) is expected to observe approximately 2000 type Ia supernovae within 
a year up to a redshift $z \sim 2$ and to improve luminosity distance 
statistics by over an order of magnitude \cite{snap}. Measurement of the 
luminosity distance $D_L(z)$ allow us to determine the Hubble parameter, 
since~ 
\cite{star98,ss99}
\begin{equation}
H(z) = \left[ \frac{d}{dz} \left( \frac{D_L(z)}{1+z} \right) \right]^{-1}.
\label{eqn:hfz}
\end{equation}

%It is therefore meaningful to ask whether a SNAP
%type experiment can be used to determine the Statefinder pair and
%place bounds on the properties of dark energy.
%
%\begin{table*}[tbh!]
%\begin{center}
%\begin{minipage}[h]{0.9\linewidth} \mbox{} \vskip -18pt
%\bigskip
%\caption{SNAP specifications for one year period of observations}
%\label{tab:SNAP}
%\begin{tabular}{ccccc}
%{Redshift Interval}  & $z=0$--$0.2$ &  $z=0.2$--$1.2$ & $z=1.2$--$1.4$
%& $z=1.4$--$1.7$   \\\hline
%{Number of SNe}  & $50$ &  $1800$ & $50$
%& $15$
%
%\end{tabular}
%\end{minipage}
%\end{center}
%\end{table*}
%
%In order to investigate this we proceed as follows.

To determine the Statefinder pair we use the following model independent
parameterization of $H(z)$: 
\beq
H^2(x)=H^2_0~\lbrack \tilde\Omega_{m0} x^3 + A + Bx+ Cx^2 \rbrack \,\,,
\label{eq:hubfit}
\eeq
where $x=1+z$ and $A+B+C= 1-\tilde\Omega_{m0}$. This form is simpler
than that used in \cite{srss00} but it is sufficient for our purpose.
It becomes exact in the case of the $\l$CDM model (i.e., dark energy
being a cosmological constant). Note that the fact that we parameterize
$H^2(z)/H_0^2$ by a 3-parameter fit means that the real $H(z)$ curve
is {\em smoothed} over some redshift interval $z\sim z_{max}/3$. 
In principle, the value of $\tilde\Omega_{m0}$ can be somewhat larger 
than the current density in CDM + baryons if dark energy has a tracker 
component having equation of state equal to that of matter at high $z$. 
However, the difference between $\tilde\Omega_{m0}$ and $\Omega_{m0}$
(if exists at all) is known to be small: $\tilde\Omega_{m0} \lleq
1.1\Omega_{m0}$. Supernova observations of $D_L$ and relations 
(\ref{eqn:hfz}) and (\ref{eq:hubfit}) can be used to determine $A, B, C$ 
and the Statefinder pair $\lbrace r,s\rbrace$, since 
\ber
r &=& 1 - \frac{(B+Cx)x}{\tilde\Omega_m x^3 + A + Bx + Cx^2}\,\,,\nonumber\\
s &=& \frac{2(B+Cx)x}{3(3A+2Bx+Cx^2)}.
\eer

In Fig. 3 we present the results obtained from 1000 random simulations 
of a SNAP-type experiment for the `mean Statefinder statistic'
\begin{eqnarray}
\bar{r} &=& \frac{1}{z_{\rm max}}\int_0^{z_{\rm max}} r(z)\,dz\,\,,\\
\bar{s} &=& \frac{1}{z_{\rm max}}\int_0^{z_{\rm max}} s(z)\,dz\,\,
\end{eqnarray}
with $z_{\rm max} = 1.7$. The simulated numbers of SNe Ia events for one 
year period of observations are taken to be 50, 1800, 50 and 15 for the 
redshift intervals (0 -- 0.2), (0.2 -- 1.2), (1.2 -- 1.4) and 
(1.4 -- 1.7) respectively. The statistical uncertainty in the magnitude 
of SNe is assumed to be constant over redshift and is given by 
$\sigma_{\rm mag} = 0.15$. Details will be presented in a companion paper~
\cite{alam}. Fig. 3 shows that a future SNAP-type experiment determining 
$\lbrace r,s\rbrace$ can easily distinguish a fiducial $\l$CDM model from 
several alternative time-dependent forms of dark energy, including
the inverse power-law quintessence potential $V \propto \phi^{-\alpha}$
with $\alpha \sim 1$ 
and the DGP brane cosmological model.

\noindent
\begin{figure}[tbh!]
\centerline{
\psfig{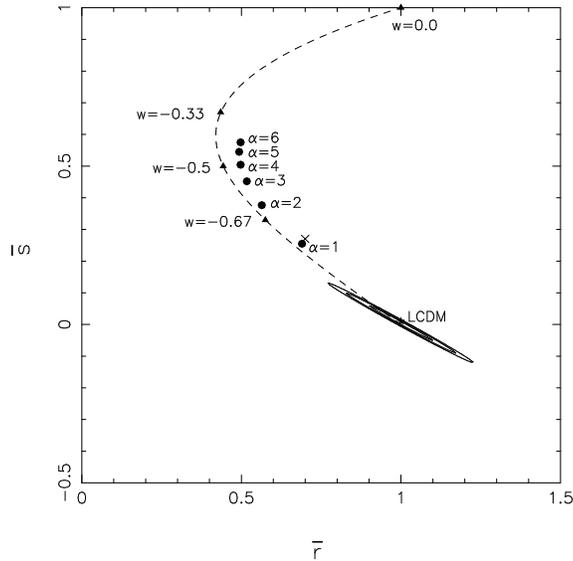} }
\caption{\small
Confidence levels at $1\sigma$, $2\sigma$, $3\sigma$ of $\rb$ and
$\sb$ computed from $1000$ random realizations of a SNAP-type experiment 
probing a $\Lambda$CDM fiducial model with $\Omega_{m0} = 0.3,~ 
\Omega_{\Lambda 0} = 0.7$. The filled circles represent the values of 
$\rb$ and $\sb$ for the quintessence potential $V(\phi) \propto 
\phi^{-\alpha}$ with $\alpha=1,2,3,4,5,6$ (bottom to top). 
The filled triangles represent quiessence with $w = -2/3, -1/2, -1/3, 0$ 
(bottom to top). The cross shows the mean Statefinder value $\rb=0.70,~
\sb=0.27$ for the DGP brane model with $H_0r_c=1.43~(\Omega_{m0}=0.3)$. 
Note that all inverse power-law models, as well as the DGP model, lie well 
outside of the three sigma contour centered around the $\l$CDM model. }
\label{fig:snap.ps}
\end{figure}

%{\it Discussion:}
%\section{Discussion}
%Observations of high redshift supernovae suggest that the Universe
%is accelerating, fueled by an unknown form of `dark energy'
%with large negative pressure. In this letter we show that
%a new and versatile cosmological parameter pair $\lbrace r,s\rbrace$
%called `Statefinder'
%can effectively distinguish between three possible forms for dark energy:
%a cosmological constant ($w = -1$), Quiessence
%($w = {\rm constant} \neq -1$) and Kinessence ($w \neq {\rm constant}$).
%The Statefinder is constructed from the scale factor and its derivatives 
%and presents a companion pair to the Hubble and deceleration  parameters
%$\lbrace H, q\rbrace$. 
%
%\medskip
%\noindent{\it Acknowledgments:}

VS acknowledges support from the ILTP program of cooperation between India 
and Russia. TDS and UA thank the UGC for providing support for this work. 
AS was partially supported by the Russian Foundation for Basic Research,
grants 02-02-16817 and 00-15-96699, and by the Research Program
"Astronomy" of the Russian Academy of Sciences.

\end{document}